\documentclass[prd,aps,11pt,superscriptaddress,nofootinbib]{revtex4}

\usepackage{color}

\usepackage{amssymb}
\usepackage{amsmath}
\usepackage{graphicx}
\usepackage{slashed}

\newcommand{\addspan}[1]{#1}

\DeclareMathOperator{\BR}{BR}
\newcommand{\GeV}{\,\text{GeV}}

\newcommand{\al}[1]{\begin{align}#1\end{align}}
\newcommand{\paren}[1]{\left(#1\right)}
\newcommand{\fn}[1]{\!\left(#1\right)}

\newcommand{\ab}[1]{\left|#1\right|}
\newcommand{\nn}{\nonumber\\}
\newcommand{\bb}{\begin{bmatrix}}
\newcommand{\eb}{\end{bmatrix}}

\newcommand{\muhat}{\hat\mu}
\newcommand{\oli}{\overline}

\begin{document}

\title{\Large Minimal dilaton model\bigskip}
\author{\large Tomohiro Abe}
\email{tomohiro_abe@tsinghua.edu.cn}
\affiliation{Institute of Modern Physics and Center for High Energy Physics, Tsinghua University, Beijing 100084, China}

\author{\large Ryuichiro Kitano}
\email{kitano@tuhep.phys.tohoku.ac.jp}
\affiliation{Department of Physics, Tohoku University, Sendai 980-8578, Japan}

\author{\large Yasufumi Konishi}
\email{konishi@krishna.th.phy.saitama-u.ac.jp}
\affiliation{Department of Physics, Saitama University, Saitama 355-8570, Japan}

\author{\large Kin-ya Oda}
\email{odakin@gauge.scphys.kyoto-u.ac.jp}
\affiliation{Department of Physics, Kyoto University, Kyoto 606-8502, Japan}

\author{\large Joe Sato}
\email{joe@phy.saitama-u.ac.jp}
\affiliation{Department of Physics, Saitama University, Saitama 355-8570, Japan}

\author{\large Shohei Sugiyama\bigskip}
\email{shohei@icrr.u-tokyo.ac.jp}
\affiliation{Institute for Cosmic Ray Research (ICRR), University of Tokyo, Kashiwa, Chiba 277-8582, Japan\bigskip\bigskip}

\begin{abstract}\normalsize
We construct a minimal calculable model of a light dilaton based on the scenario where only top and Higgs sectors are involved in a quasiconformal dynamics. The model consistently accommodates the electroweak precision tests even when the Higgs boson is very heavy, thereby allowing one to consider the possibility that the particle at around 125\,GeV, discovered at the LHC experiments, is identified as the light dilaton rather than the Higgs boson. We find that the current LHC data allow distinct parameter regions where the observed particle is either mostly the Higgs boson or the dilaton.
\end{abstract}

\begin{flushright}
KUNS-2417,
STUPP-12-211,
TU-919
\end{flushright}

\maketitle

\newpage

\section{Introduction}

It has been reported recently that both the ATLAS and the CMS
experiments observed resonances in the $\gamma \gamma$, $ZZ$, and $WW$
channels at around 125\,GeV~\cite{obs:2012gk, obs:2012gu}. The resonance
can naturally be interpreted as a signal of the Higgs boson in the
Standard Model (SM).
The mass range suggests that the Higgs sector of the SM is
fairly weakly coupled, perfectly consistent with the precision
measurements at the LEP experiments.

The observation of the light Higgs boson excludes various models of
electroweak symmetry breaking via strong dynamics. For example,
typical technicolor models~\cite{TC} and their proposed effective
descriptions 
such as the Higgsless model~\cite{Csaki:2003dt} are in trouble with the
weakly coupled descriptions of the Higgs sector.

There is, however, a logical possibility that the excess is not due to
the Higgs boson but rather a Nambu-Goldstone boson associated with an
approximate scale invariance in the dynamical sector. In Ref.~\cite{WTC_TD},
it has been proposed that the dilaton in the walking technicolor~\addspan{\cite{WTC_TD_original}} may explain
the signals, \addspan{see also Ref.~\cite{light_dilaton} for an effective theory description of dilaton systems.}
(There is an attempt to identify a dilaton as the Higgs~\cite{Higgs as PNGB}.)
In Ref.~\cite{radion in RS as 125GeV},
the radion in the Randall-Sundrum model has
been discussed as a possible candidate, see also~Ref.~\cite{radion further development}.
%

It has been reported that the excess of events at around 126.5 (125) GeV at ATLAS (CMS) in the diphoton final state is enhanced from the expectation of the SM by a factor $1.8\pm0.5$ ($1.6\pm0.4$)~\cite{obs:2012gk, obs:2012gu}.
Though this tendency is not to the level of
significance, it is noteworthy that the above stated models can
naturally account for the trend.

If the excesses at around 125\,GeV are due to such a non-Higgs particle,
one needs to consider a consistency with the constraints on the
Peskin-Takeuchi $S$ and $T$ parameters~\cite{PeskinTakeuchi}.
This non-Higgs scenario requires the real Higgs boson (if it exists) to be
heavier than $\sim 600$\,GeV in order to be consistent with the LHC
data, and such a heavy Higgs boson does not give a good fit to the
precision measurements. It is required to have some other contributions
to the $S$ and $T$ parameters.
Since the models based on (unknown) technicolor theories or an extra dimension have no predictability or only weak predictability on the $S$ and $T$ parameters, it is difficult to judge if such models are really viable.

In this paper, we construct an effective (minimal) model of such a framework
and see if there is a viable parameter region. The model consists of a
massive vectorlike top partner fermion. The top partner mass $M$ represents
a mass gap in the dynamical sector, to which the dilaton naturally couples 
in order to recover the scale invariance: 
$M \to M
e^{-\phi/f}$.
The coupling in turn provides interactions between the dilaton $\phi$
and the photons/gluons through loop diagrams of the top partners,
explaining the excesses at the ATLAS and CMS.
The top partner can also contribute to the Peskin-Takeuchi $S$, $T$ parameters, which
turns out to be a good direction to come back to the allowed region,
canceling the heavy SM Higgs boson contribution to the $T$ parameter.

We discuss whether such a dilaton explanation of the excesses is viable
and how such a scenario can be discriminated by measuring various cross
sections and decay branching ratios at the LHC experiments.
The minimal model we study below catches essential features of
dilaton/radion models, and the model parameters we discuss can easily be
translated into those in other models.

In Sec.~\ref{sec:model}, we present the minimal dilaton model as an effective renormalizable theory equipped with a linearized dilaton field $S$ and the vectorlike top partner $T$~\footnote{
There would be no room for confusion from the notational abuse that the linearized dilaton field~$S$ and the top partner $T$ should not be misunderstood as the Peskin-Takeuchi parameters which appear with superscripts only in Eqs.~\eqref{S-top equation}--\eqref{eq:T-scalar}.}.
In Sec.~\ref{Dilaton at the LHC}, we show signal strengths for all the Higgs decay final states and give constraints on model parameters, namely a Higgs-dilaton mixing angle $\theta_H$ and a dilaton decay constant $f$, which is nothing but a vacuum expectation value of $S$.
In Sec.~\ref{top_constraints}, constraints from the electroweak precision measurements are examined  on the top partner sector, namely on the left-handed mixing $\theta_L$ between the top and its partner $T$ and on the heavier $t'$ quark mass.
The last section is devoted to summary and discussions.

\section{The minimal dilaton model}
\label{sec:model}

The dilaton field, defined as the Nambu-Goldstone particle associated
with an approximate scale invariance of the theory, can be produced
through the gluon fusion process at the LHC and can decay into two photons
through the following effective operators:
\begin{eqnarray}
 \phi G^a_{\mu \nu} G^{a \mu \nu},\ \ \ 
 \phi F_{\mu \nu} F^{\mu \nu},
\label{eq:operators}
\end{eqnarray}
where $\phi$, $G^a_{\mu \nu}$, and $F_{\mu \nu}$ are the dilaton, the
gluon field strength, and the photon field strength, respectively.
These effective operators are generated if there is a colored and
charged field which obtains mass through the spontaneous breaking of the
approximate scale invariance.
For example, in an approximately scale invariant technicolor theory, one
can expect such operators to appear at low energy if there are colored
and charged techniquarks.
Also, there can be a dual hadronic description of such a theory where
the approximate scale invariance is nonlinearly realized. An example is
a model with a warped extra dimension where $\phi$ represents the
radius of the extra dimension.
In either example, predictions to the effective couplings and also
constraints from the electroweak precision measurements are pretty model
dependent, and moreover, there are often technical difficulties in the
estimations due to large nonperturbative effects or incalculable
corrections from the cutoff scale physics.

We, therefore, consider an effective minimal model of the dilaton, based
on a weakly coupled renormalizable theory.
The model allows us to perform explicit computations of the Peskin-Takeuchi $S$, $T$
parameters and also production/decay processes of the dilaton at the LHC.
This exercise not only provides us with a sense of how such a model is
constrained, but is also practically useful since the obtained allowed
range of parameters can easily be translated into other calculable
models.

The operators in Eq.~\eqref{eq:operators} are obtained by integrating
out a field which is colored and charged.
We choose the field to have the same quantum numbers as the right-handed
top quark.
This is somewhat a natural choice. When we consider the origin of the
large top-quark mass, one may need to assume that the top quark is
(semi) strongly coupled to a dynamical sector, such as in the
topcolor~\cite{Hill:1991at} or the top seesaw models~\cite{TopSeeSaw}.
It is then reasonable to
assume an existence of a resonance with the same quantum number as the
top quarks.
The resonance can decay into a bottom quark and a $W$ boson, and thus
does not have a problem with a exotic stable state.
As a minimal choice, we consider a vectorlike top partner with the same gauge quantum number as the right-handed $SU(2)_L$ singlet top quark rather than the left-handed doublet that also includes bottom quark partner.

We write down the following Lagrangian for the dilaton system:
\begin{eqnarray}
 {\cal L} = {\cal L}_{\rm SM}
- {e^{-2 \phi /f } \over 2 }
\partial_\mu \phi \partial^\mu \phi
- \oli{T}\left(\slashed{D}+M e^{- \phi/f} \right)T
- \left[
y' \oli{T_R}(q_{3L}\cdot H)
+ \text{h.c.} \right]
- V(\phi, H),
	\label{nonlinear_model}
\end{eqnarray}
where $T$ is the heavy vectorlike top partner representing the 
resonance, $q_{3L}$ is the left-handed (top and bottom) quark doublet, and $H$ is the SM Higgs doublet field.
The Lagrangian ${\cal L}_{\rm SM}$ is the Standard Model, and the term
with a coupling constant $y'$ provides a mixing between the top
quark and $T$.
The Lagrangian has a nonlinearly realized scale invariance except for
the scalar potential term $V(\phi,H)$ which contains terms with small
explicit breaking of the scale invariance.
The potential terms provide mass terms for $\phi$ as well as a mixing
between $\phi$ and the Higgs boson.
We choose the origin of the field $\phi$ so that $\langle \phi \rangle =
0$. A mass term of $\oli{u_{3R}} T_L$, with $u_{3R}$ being
the right-handed top quark, can be eliminated by an appropriate field
redefinition; see Appendix~\ref{top_mixing_section}.
It may be interesting to consider this model as the low-energy effective
theory of the top condensation model~\cite{TopCondensation},
where the coupling $y'$ and
the quartic coupling constant of the Higgs field blows up
simultaneously at a high-energy scale. We do not impose such a
constraint in this paper in order to leave the discussion general.

\subsection{Linearized model}

By a field redefinition,
\begin{eqnarray}
 S = f e^{-\phi / f},
\end{eqnarray}
the Lagrangian given in Eq.~\eqref{nonlinear_model} is equivalent to
\begin{eqnarray}
 {\cal L} = {\cal L}_{\rm SM}
- {1 \over 2 }
\partial_\mu S \partial^\mu S
- \oli{ T}\left(\slashed{D}+{M \over f} S\right)T
- \left[
y' \oli{T}(q_{3L}\cdot H)
+ \text{h.c.} \right]
- \tilde V(S, H).
\label{linear_realization}
\end{eqnarray}
The scale invariance is now linearly realized.
The potential $\tilde V (S,H)$ should be arranged so that $\langle S
\rangle = f$ and $\langle H^0 \rangle = v/\sqrt 2$.
The explicit form of $\tilde V$ is shown in Appendix~\ref{linear_realization_potential} for completeness, though we do not need to specify it for the following analysis as we will be discussing with physical quantities such as the masses and mixings.
We propose this Lagrangian as a minimal effective description of an approximately scale invariant theory of (dynamical) electroweak symmetry breaking involving the only top and Higgs sectors.

If the mixing between the dilaton and the Higgs boson is small and if
the dilaton is the one which explains the observed resonance at the LHC,
the Higgs boson must be heavier than about $600$\,GeV in order to be
consistent with the Higgs boson searches at the LHC. With such a heavy
Higgs boson, constraints from the Peskin-Takeuchi $S$, $T$ parameters require a new
contribution to come back to the ellipse in the $S$-$T$ plane. As we will
see shortly, such a contribution is already there in this model since
the loop diagrams of the $t^\prime$ field can push $S$ and $T$ parameters towards
the right direction.
This Lagrangian, therefore, provides a compact realistic model of the light
dilaton, which can be used for LHC studies.

\subsection{Model parameters}
This model has new parameters in addition to the Standard Model ones. We
here list them and their definitions:
\begin{itemize}
 \item $f$

This is the decay constant of the dilaton. The size of $f$ controls the
       strength of the coupling of the dilaton $\phi$ to photons and gluons and to all the fields involved in the quasiconformal dynamics. For later use, we define a dimensionless quantity,
\begin{eqnarray}
\eta = {v \over f} N_{T},
\label{xi_eq}
\end{eqnarray}
where $v = \sqrt 2 \langle H^0 \rangle = 246$\,GeV, and $N_{T}$ is
       the number of the $T$ fields.  
This parameter appears when we discuss the production and decay of
       $\phi$.
If necessary, one can obtain a large value of $\eta$ with a large number of $N_T$ without requiring too small $f$. 
For the minimal model
       $N_{T} = 1$ which we will assume hereafter.

 \item $m_s$, $m_h$, and $\theta_H$

These are masses ($m_s < m_h$) and mixing of the scalar fields. We take
       lighter mass eigenstate to explain the LHC excesses, {\it i.e.,}
\begin{eqnarray}
 m_s \simeq 125\,{\rm GeV}.
\end{eqnarray}
The Higgs-dilaton mixing angle is defined as
\al{
 S	&=	f + s \cos \theta_H - h \sin \theta_H,	&
 H^0
 	&=	{v+s\sin\theta_H+h\cos\theta_H\over\sqrt{2}},
\label{eq:thetaH}
}
where $s$ and $h$ are the lighter and the heavier mass eigenstates,
       respectively.
We impose
\begin{eqnarray}
 m_h > 600~{\rm GeV},
\end{eqnarray}
to be consistent with the data from LHC. When the mixing angle is so
       large that the lighter one is almost the Standard Model 
       Higgs boson, it is not necessary to impose the above constraint
       for a large $f$. Since we are particularly interested in a small
       mixing region, we always impose the above constraint in the
       following analysis.

 \item $m_{t'}$ and $\theta_L$

These are the $T$ mass and the left-handed mixing of the top
       sector. The mass matrix for the top quark and its partner,
\al{
\mathcal M_t
	&=	\bb
		y_t v / \sqrt 2 & y' v / \sqrt 2 \\
		0 & M
		\eb,
		\label{eq:massmat}
}
are diagonalized as
\al{
\bb
\cos\theta_L	&	-\sin\theta_L\\
\sin\theta_L	&	\cos\theta_L
\eb
\mathcal M_t\mathcal M_t^\dagger
\bb
\cos\theta_L	&	\sin\theta_L\\
-\sin\theta_L	&	\cos\theta_L
\eb
	&=	\bb
		m_t^2	&	0\\
		0	&	m_{t'}^2
		\eb,
}
where $m_t$ is the top quark mass.
One can trade the new parameters
       $y'$ and $M$ by the observables $m_{t'}$ and $\theta_L$, while $y_t$
       should be adjusted to reproduce the top quark mass.
See Appendix~\ref{top_mixing_section} for more detailed discussion.
In Sec.~\ref{top_constraints}, we verify constraints on the parameters $\theta_L$ and $m_{t'}$ from the electroweak precision data.

\end{itemize}

\subsection{Coupling to the Standard Model fields}
\label{theoretical_explanation}

The loop diagrams of $T$ generate the effective couplings in
Eq.~\eqref{eq:operators}. In the limit of $2 m_{t'} \gg m_s$, the
effective couplings for the production and decays of $s$ are approximated
by momentum independent pieces which are insensitive to $m_{t'}$
or $\theta_L$. 
This behavior can most easily be understood by identifying the effective coupling
as the dilaton/Higgs dependence of the running gauge coupling constants:
\begin{eqnarray}
 {\cal L}_{\rm eff} = - {1 \over 4 g_s^2} G^a_{\mu \nu} G^{a \mu \nu},
\label{eq:eff}
\end{eqnarray}
where we take the gluon as an example. The running coupling constant at
a low-energy scale is given by
\begin{eqnarray}
 {1 \over g_s^2 (\mu)} = {1 \over g_s^2 (\Lambda)}
- {2 (b_{\rm SM} + \Delta b) \over (4\pi)^2} \log {m_{t'} \over
 \Lambda}
- {2 b_{\rm SM} \over (4\pi)^2} \log {m_t \over m_{t'}}
- {2 (b_{\rm SM} - \Delta b) \over (4\pi)^2} \log {\mu \over m_t},
\end{eqnarray}
at one-loop level, where $b_{\rm SM} = -7$ and $\Delta b = 2/3$. This
can be rearranged to
\begin{eqnarray}
 {1 \over g_s^2 (\mu)}
&=& {1 \over g_s^2 (\Lambda)}
- {2 (b_{\rm SM} - \Delta b) \over (4\pi)^2} \log {\mu \over \Lambda}
- {2 \Delta b \over (4\pi)^2} 
\log {(y_t \langle H^0 \rangle) (M \langle S \rangle /f) \over \Lambda^2},
\end{eqnarray}
where we have used the fact that 
\begin{eqnarray}
 m_t m_{t'} 
= y_t \langle H^0 \rangle \, {M \langle S \rangle \over f},
\end{eqnarray}
which is derived from the mass matrix in Eq.~\eqref{eq:massmat}. 
By recovering the field fluctuations $\langle H^0 \rangle,\langle S
\rangle$ $\to H^0, S$ and considering the mixing factors in
Eq.~\eqref{eq:thetaH}, we obtain the effective coupling of $s$ to gluons
from Eq.~\eqref{eq:eff}
as
\begin{eqnarray}
 {\cal L}_{\rm eff}^{(sgg)} = {1 \over 4} \, {g_s^2 \over (4\pi)^2}
{2 \over v}\,{2\over3}
\left(
\eta \cos\theta_H + \sin \theta_H
\right) 
s
G^a_{\mu \nu}
G^{a \mu \nu},
\end{eqnarray}
where we have canonically normalized the kinetic term of the gluon.
The terms proportional to $\sin\theta_H$ and $\eta \cos\theta_H$ are contributions from the SM top and its partner, respectively.
We can explicitly see that the coupling is independent of $m_{t'}$
or $\theta_L$.
For the $s$ to photon coupling, we need to include a loop of the $W$
bosons. The result is
\begin{eqnarray}
  {\cal L}_{\rm eff}^{(s \gamma \gamma)} 
= {1 \over 4} \, {e^2 \over (4\pi)^2}
{2 \over v} 
\left(
{4 \over 3} N_c Q_t^2 \eta \cos\theta_H  + A_\text{SM} \sin \theta_H
\right) 
s
F_{\mu \nu}
F^{\mu \nu},
\end{eqnarray}
where $N_c=3$ and $Q_t = 2/3$ are the color factor and the top quark charge, respectively, and the explicit form of the loop factor $A_\text{SM} \simeq - 6.5$ can be found e.g.\ in  (2.45) in Ref.~\cite{Djouadi:2005gi}.
The first term in the parentheses is the contribution from the top partner loop whereas the second is from the SM top and $W$ ones.

The particle $s$ can also couple to the $W$ and $Z$ bosons and the
fermions through the $\theta_H$ mixing. The couplings are simply given
by those of the SM Higgs boson times a factor of $\sin
\theta_H$.

Note here that the model is not the same as the Higgs-dilaton model studied in
Refs.~\cite{Giardino:2012ww,Giardino:2012dp,Carmi:2012in}, where the coupling between the dilaton and the Standard
Model fields are assumed to have the form:
\begin{eqnarray}
 {\cal L}_{\rm int} = {\phi \over f} T^{\mu}_{\ \mu}.
\label{eq:phiT}
\end{eqnarray}
Here, $T^{\mu}_{\ \mu}$ is the trace of the energy-momentum tensor of
the Standard Model.
Through this term, the dilaton directly couples to the violation of the
scale invariance in the Standard Model, i.e., to the $W$, $Z$ bosons
and fermions with strength proportional to their masses.
Also, the couplings to the photons and gluons are proportional to the
beta functions.
The effective interaction term in Eq.~\eqref{eq:phiT} is generated at
low energy if the whole Standard Model sector is a part of the scale
invariant theory in the UV; all the gauge bosons and fermions are
composite particles.

In contrast, we take a more conservative picture that the Standard Model
except for the top/Higgs sector is a spectator of the dynamics, and thus
the dilaton couples to the $W$, $Z$ bosons and fermions only through
the mixing with the Higgs fields. The couplings to the gluons and
photons are generated only through the loops of $t$ and $t^\prime$.
Due to these different origin of the couplings between two models, the
production and decay properties are quite different.
Indeed, we will see that our model can give better fit to the LHC data compared to the SM Higgs boson, while Refs.~\cite{Giardino:2012ww,Giardino:2012dp,Carmi:2012in}
have reported that the dilaton scenario based on Eq.~\eqref{eq:phiT} is
rather disfavored.

In terms of the parameters in Refs.~\cite{Giardino:2012ww,Giardino:2012dp,Carmi:2012in}, our model corresponds to
\al{
c_V	=	c_F
	&=	\sin\theta_H, &
c_t
	&=
		\cos^2\theta_L\sin\theta_H+\eta\sin^2\theta_L\cos\theta_H, &
}
\al{
c_g	&=	\eta\cos\theta_H+\sin\theta_H,	&
c_\gamma
	&=	\eta A_{t'}\cos\theta_H+A_\text{SM}\sin\theta_H,	&
c_\text{inv}	&=	0,
}
where $F$ stands for all the SM fermions except the top quark.
(The parameter $c_X$ for a production/decay process $X$ is written as $\kappa_X$ in a recent analysis by ATLAS~\cite{ATLAS coupling}. Our notation in the forthcoming Eqs.~\eqref{GF production ratio}--\eqref{decay ratios} reads $R_X=c_X^2=\kappa_X^2$.)
It is worth
noting that a negative value for $c_F$ can be easily obtained in our
model, which tends to be more favored than the SM value $c_F=1$ in order
to suppress the $c_g$ coupling while keeping $c_\gamma$
large~\cite{Giardino:2012ww,Giardino:2012dp,Carmi:2012in}.

\section{Dilaton at the LHC}\label{Dilaton at the LHC}

As we discussed in Sec. \ref{sec:model}, there are two mass
eigenstates in the scalar sector, $s$ and $h$, and their couplings are
determined by two parameters, $\eta$ and $\theta_H$ in
Eqs.~\eqref{xi_eq} and \eqref{eq:thetaH}. In the small $\theta_H$ region,
$s$ is dilatonlike as we can see from Eq.~\eqref{eq:thetaH}.
In the following, we assume the lighter mass eigenstate $s$ to be
around 125\,GeV and study the production and decays of $s$ at the LHC.

\subsection{Production}

\begin{figure}
\begin{center}
\hfill
\includegraphics[width=.4\textwidth]{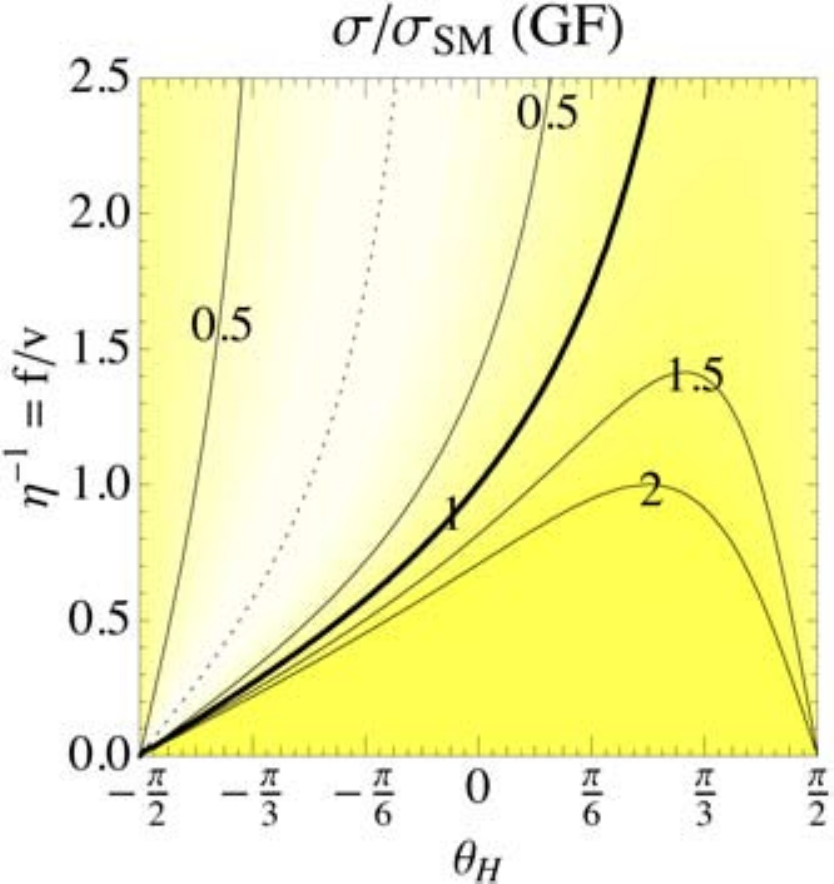}\hfill
\includegraphics[width=.4\textwidth]{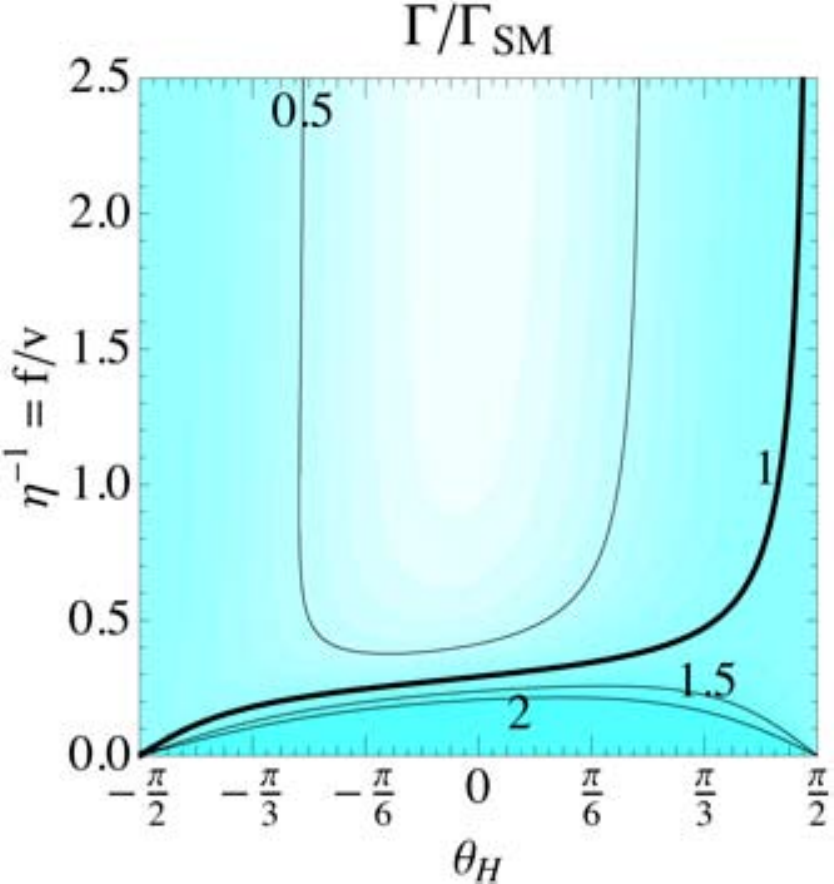}\hfill\mbox{}
\\
\caption{Ratio to the SM for the dominant GF production cross section
 given in Eq.~\eqref{GF production ratio} (left) and to the total decay width given in
 Eq.~\eqref{decay_width} (right). Contours 0, 0.5, 1, 1.5, and 2 are
 drawn, with 0 and 1 being dotted and thick lines, respectively. A denser
 region gives larger value, with density changing for each increase of
 the ratio by 0.1 from 0 to 2. Both sides $\theta_H=\pm \pi/2$
 correspond to the SM.} 
\label{total_ratio}
\end{center}
\end{figure}

As one can see from the above discussion, the $s$ particle has
suppressed couplings to $W$, $Z$ and fermions, and either enhanced or
suppressed couplings to $\gamma$ and the gluon compared to the Higgs
boson in the Standard Model. The production cross section of $s$
compared to that of the SM Higgs boson (at the same mass as $s$) through a process $X$,
\al{
R_X	&:=	{\sigma_X\over\sigma_X^\text{SM}},
		\label{production_ratio}
}
is given by
\al{
R_\text{GF}
	&=	\paren{\eta\cos\theta_H+\sin\theta_H}^2,	
		\label{GF production ratio}\\
R_\text{VBF}
	=	R_\text{VH}
	&=	\sin^2\theta_H,	\\
R_\text{ttH}
	&=	\paren{\cos^2\theta_L\sin\theta_H+\eta\sin^2\theta_L\cos\theta_H}^2,
		\label{ttH production ratio}
}
for the gluon fusion (GF), the vector boson fusion (VBF), the
Higgs-strahlung (VH), and the associated production with a $t\bar t$ pair (ttH), respectively; 
see
 e.g.\ Ref.~\cite{Djouadi:2005gi} for a review. Here and hereafter $V$
 ($VV$) denotes either $W$ or $Z$ ($WW$ or $ZZ$). Note that the SM cross
 section in the denominator in Eq.~\eqref{production_ratio} is evaluated
 at $m_h\simeq125$\,GeV for comparison to the experimental data, while
 its value in our model is $m_h\gtrsim 600$\,GeV. As said above, $\eta$
 appearing in Eq.~\eqref{GF production ratio} is not a ratio of the Yukawa couplings
 but that of  the vacuum expectation values which can also be checked by
 a direct loop computation in the linearized model given in
 Eq.~\eqref{linear_realization}.
 We plot $R_\text{GF}$ in the left panel of Fig.~\ref{total_ratio}. 
 The ttH ratio~\eqref{ttH production ratio} reduces to $R_\text{ttH}\to\sin^2\theta_H$ in the small top mixing limit $\theta_L\ll1$, which we will assume hereafter. The validity of this approximation will be confirmed in Sec.~\ref{top_constraints}. We note that the ttH process gives negligible contribution to our diphoton analysis in Eq.~\eqref{signal_strength}, with $\varepsilon^i_\text{ttH}$ being at most 4\%, and that we include it just for completeness.

\subsection{Decay}\label{decay_subsection}
For each decay process $s\to X$, we define the decay width ratio to that of the SM Higgs at 125\,GeV,
\al{
R(s\to X)
	&=	{\Gamma_{s\to X}\over\Gamma_{h\to X}^\text{SM}}.
}
The minimal dilaton model predicts
\al{
R(s\to\text{others})
	&=	\sin^2\theta_H,	\\
R(s\to gg)
	&=	\paren{\eta\cos\theta_H+\sin\theta_H}^2,	\\
R(s\to \gamma\gamma)
	&=	\paren{\eta{A_{t'}\over A_\text{SM}}\cos\theta_H+\sin\theta_H}^2,
		\label{decay ratios}
}
where the subscript ``others'' denotes the tree-level processes $bb,VV,\tau\tau,cc$, etc.\  and 
\al{
A_{t'}
	&:=	N_cQ_{t'}^2\,A_{1\over2}\fn{m_s^2\over4m_{t'}^2}
	\simeq {16\over9},
}
with the loop function $A_{1\over2}$ given in Eq.~(2.46) in
Ref.~\cite{Djouadi:2005gi}. 
The ratio for the total decay width is 
\al{
R(s\to\text{all})
	&=
			{\Gamma_{s\to \text{others}}
			+\Gamma_{s\to gg}
			+\Gamma_{s\to \gamma\gamma}
			\over
			\Gamma_{h\to\text{all}}^\text{SM}
			}\nn
	&=	\BR_\text{others}^\text{SM}\,
			\sin^2\theta_H
		+\BR_{gg}^\text{SM}\,\paren{\eta\cos\theta_H+\sin\theta_H}^2
		+\BR_{\gamma\gamma}^\text{SM}\,
			\paren{\eta{A_{t'}\over A_\text{SM}}\cos\theta_H+\sin\theta_H}^2,
			\label{decay_width}
}
where at around 125\,GeV, branching ratios in the SM are given e.g.\ in Ref.~{\cite{Giardino:2012ww}}
\al{
\BR_\text{others}^\text{SM}
	&=	91.3\%,	&
\BR_{gg}^\text{SM}
	&=	8.5\%,	&
\BR_{\gamma\gamma}^\text{SM}
	&=	0.2\%.
}
We plot 
$R(s\to\text{all})$
in the right panel of Fig.~\ref{total_ratio}.

\subsection{Dilaton vs SM Higgs signal strengths}
Both ATLAS and CMS experiments discovered a new particle
 at around 125\,GeV in the diphoton~\cite{CMS_diphoton,ATLAS_diphoton}, $ZZ\to4l$~\cite{CMS_ZZ4l,ATLAS_ZZ4l}, and $WW\to l\nu l\nu$~\cite{CMS_WWlnln,ATLAS_WWlnln} channels. The obtained data for each channel are translated into the signal strength, which is an expected production cross section for a particle that decays the same as in the SM Higgs at the same mass. We constrain the model parameters $\theta_H$ and $\eta^{-1}=f/v$ from these three channels.

The minimal dilaton model predicts different production cross sections
between GF and VBF/VH/ttH processes. In $H\to\gamma\gamma$ search,
composition of these production channels differs category by category
and are summarized in Table 2 in Ref.~\cite{obs:2012gu} for CMS and in
Table~6 in Ref.~\cite{ATLAS_diphoton} for ATLAS. 
We define $\varepsilon^i_X$ as the proportion of the production process
$X$ within a category $i$.
Note that $\sum_X\varepsilon^i_X=1$ by definition for each category
$i$, where a summation over $X$ is always understood as for all the relevant production channels: GF, VBF, VH, and ttH.
GF is the dominant production process
and satisfies $\varepsilon^i_\text{GF}\lesssim 90\%$
in production processes other than dijet category. 
In the dijet category, the dominant production process is VBF,
and $\varepsilon_\text{VBF}\lesssim 70\%$.

When acceptance of a production channel $X$ for a category $i$ is $a^i_X$, the estimated value of a signal fraction under the given set of cuts $i$ becomes
\al{
\varepsilon^i_X	&=	{a^i_X\sigma^\text{SM}_X\over\sum_Ya^i_Y\sigma^\text{SM}_Y},
	\label{varepsilon_eq}
}
where $\sigma^\text{SM}_X$ is the Higgs production cross section in the SM through the channel $X$.
Given $\{\varepsilon^i_X\}$, we can compute the signal strength under the imposed cuts for each category $i$
\al{
\muhat_i(h\to \gamma\gamma)
	&=	{\sum_Xa^i_X\,\sigma_X\over
			\sum_Y a^i_Y\,\sigma_Y^\text{SM}}\,{\BR(s\to \gamma\gamma)\over \BR(h\to\gamma\gamma)_\text{SM}}
	=	\sum_X\varepsilon^i_X R_X
				\,{R(s\to \gamma\gamma)\over R(s\to\text{all})}.
				\label{signal_strength}
}
We have assumed that the acceptance $a^i_X$ under the category $i$ does not change from that of the SM for each production channel $X$.

\begin{figure}
\begin{center}
\hfill
\includegraphics[width=.4\textwidth]{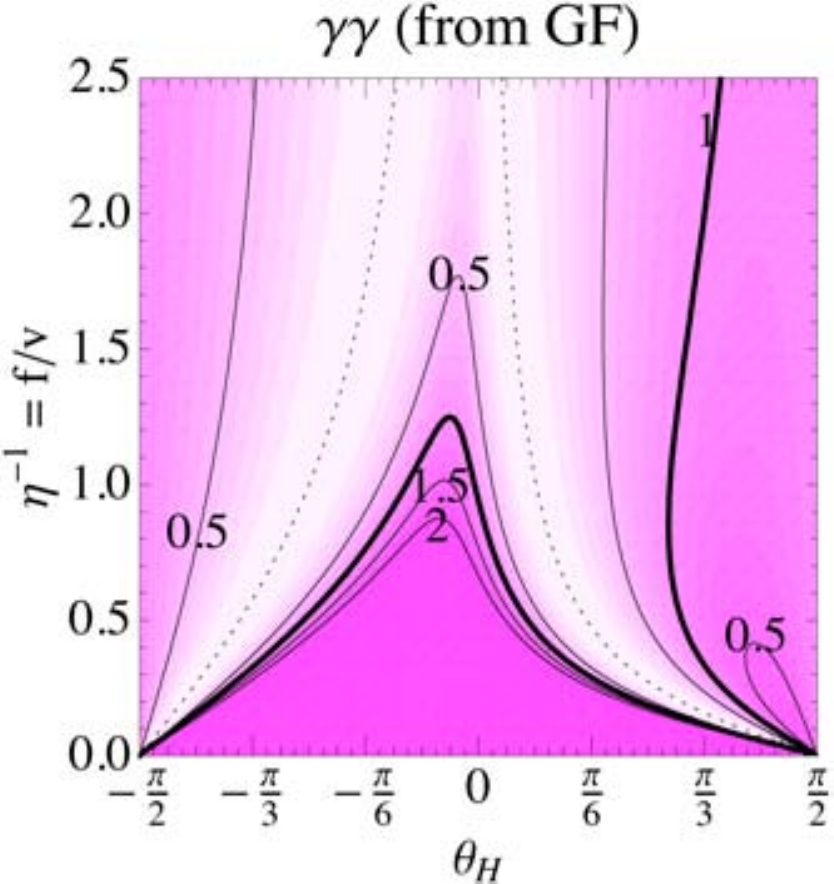}\hfill
\includegraphics[width=.4\textwidth]{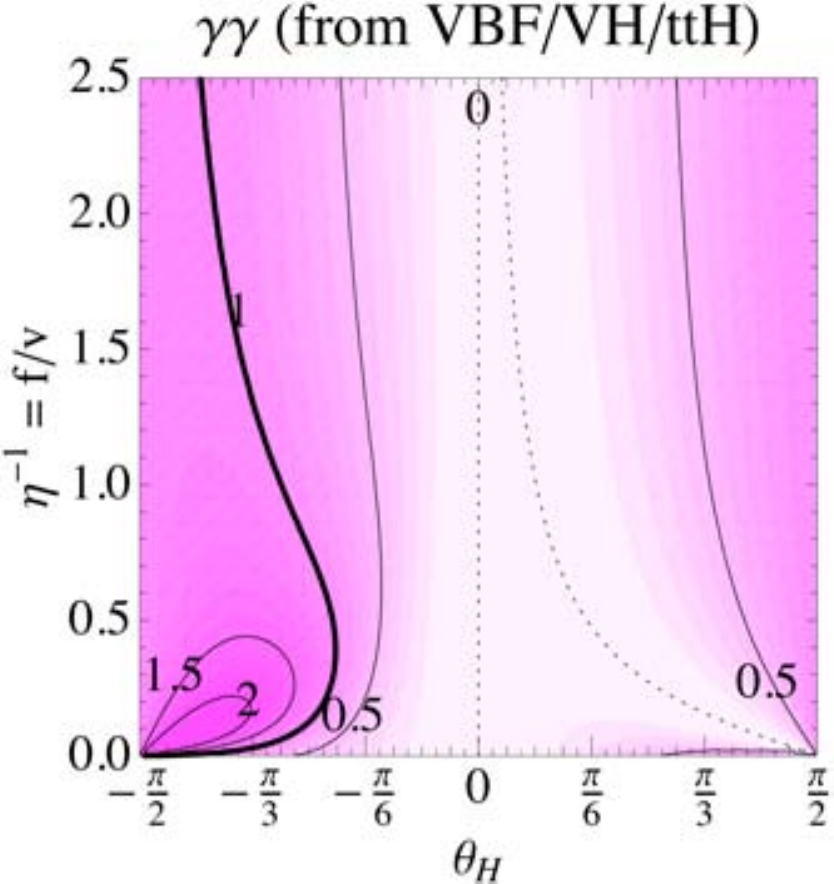}\hfill\mbox{}
\\
\caption{Diphoton $s\to\gamma\gamma$ signal strength $\muhat$ when the production is purely from the GF (VBF/VH/ttH) process in the left (right) panel. Drawn the same as in Fig.~\ref{total_ratio}. }
\label{diphoton}
\end{center}
\end{figure}

\begin{figure}
\begin{center}
\hfill
\includegraphics[width=.4\textwidth]{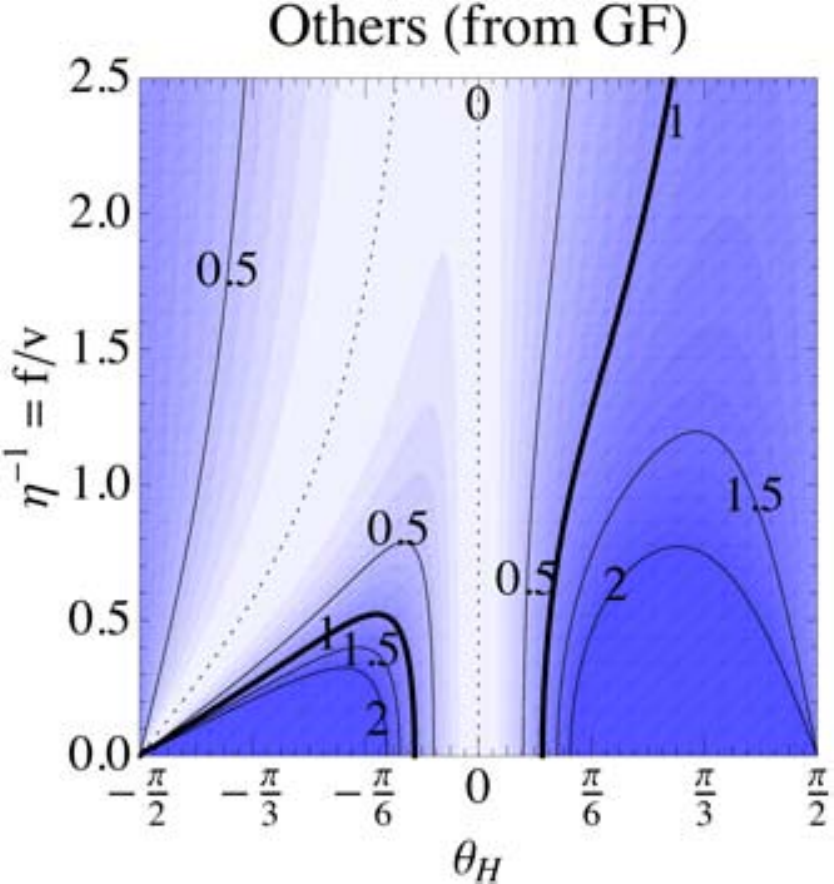}\hfill
\includegraphics[width=.4\textwidth]{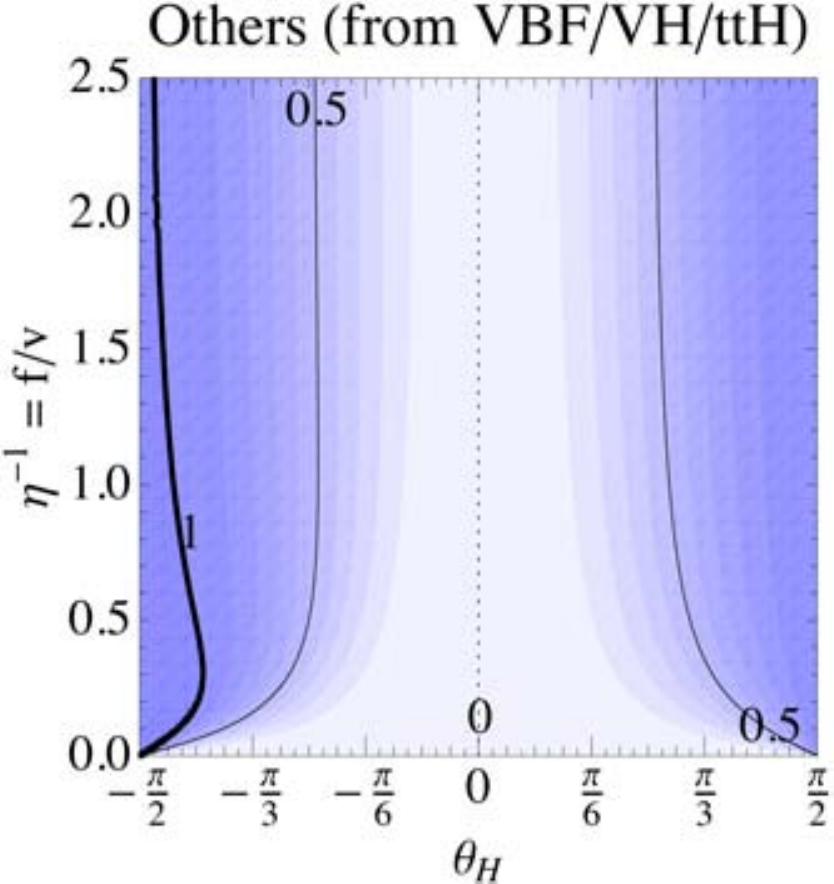}\hfill\mbox{}
\\
\caption{Signal strength $\muhat$ for processes other than diphoton and digluon, namely for final states $s\to ZZ$, $WW$, $\tau\tau$, $bb$, etc. Drawn the same as in Fig.~\ref{diphoton}.}
\label{others}
\end{center}
\end{figure}

\begin{figure}
\begin{center}
\hfill
\includegraphics[width=.4\textwidth]{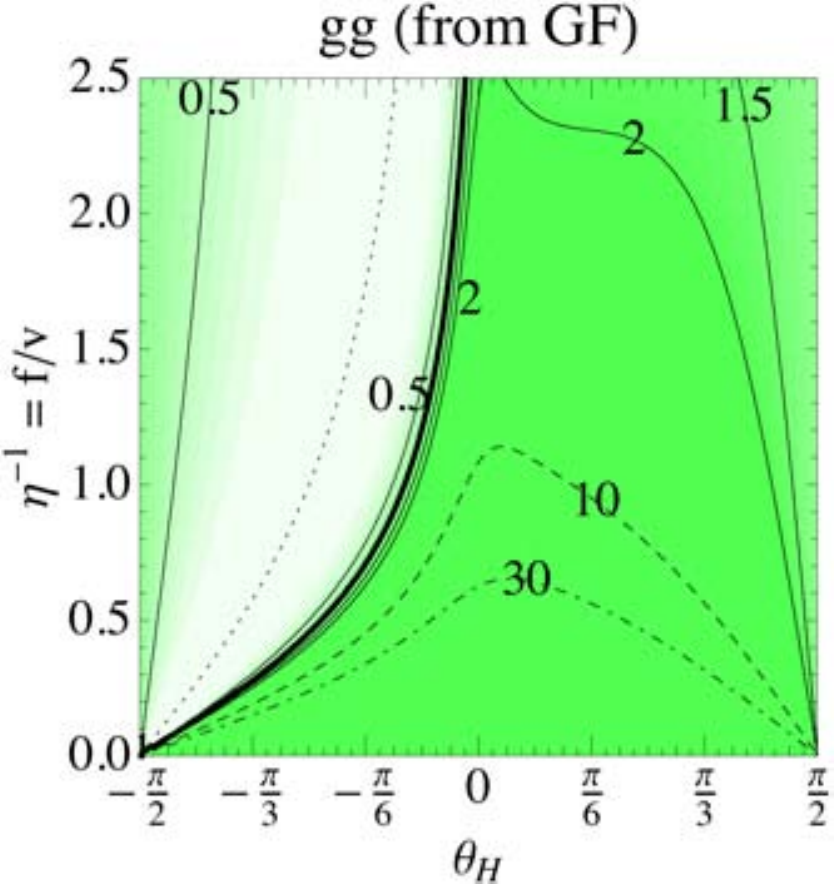}\hfill
\includegraphics[width=.4\textwidth]{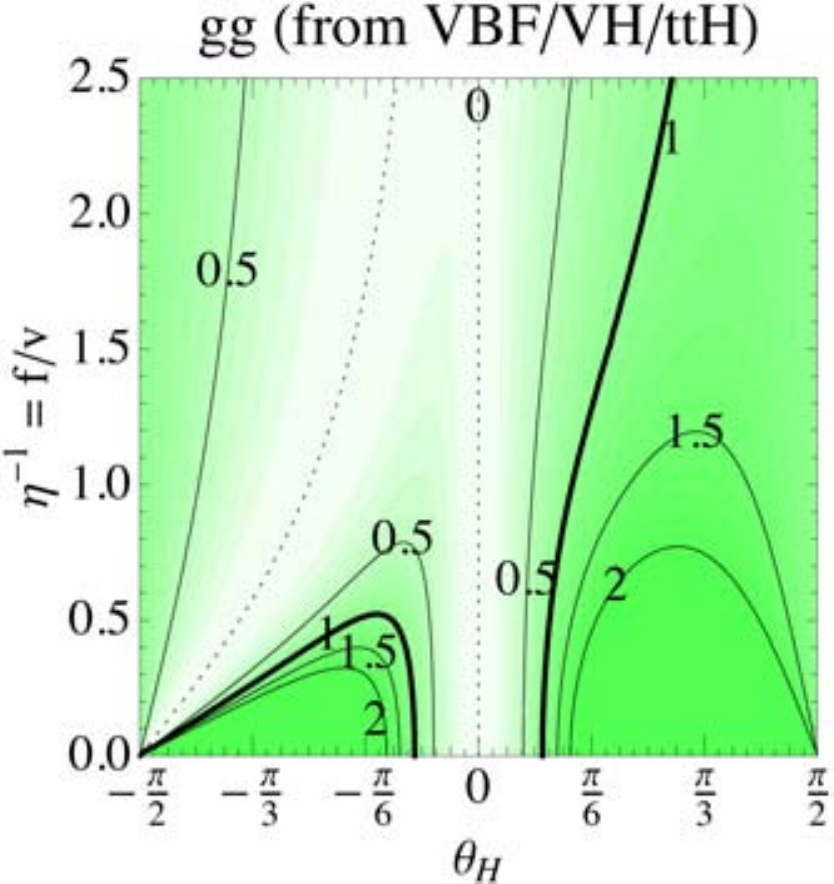}\hfill\mbox{}
\\
\caption{Signal strength $\muhat$ for the digluon $s\to gg$ process. Drawn the same as in Fig.~\ref{diphoton}. Dashed and dot-dashed contours are added for $\hat\mu=10$ and 30, respectively, in the left panel.}
\label{digluon}
\end{center}
\end{figure}

As an illustration, we plot each contribution from the initial state $X$ in the signal strength $\hat\mu(s\to F)$:
\al{
\hat\mu_X(s\to F)
	&=	R_X{R(s\to F)\over R(s\to\text{all})},
}
where explicit form of all the rates in right-hand side has already been presented in Sec.~\ref{decay_subsection}.
Figures~\ref{diphoton} and \ref{others} are, respectively, for the diphoton final state and for the others than diphoton and digluon ones. Though it is hardly observable at the LHC, we also plot the signal strength for digluon final states in Fig.~\ref{digluon} for
completeness. For real experimental data under a given set of cuts $i$, signal
strength becomes a mixture of those from GF and VBF,VH,ttH processes
shown in the left and right panels, respectively, with coefficients
$\varepsilon^i_X$ ($X=\text{GF, VBF, VH, and ttH}$) being multiplied as in
Eq.~\eqref{signal_strength}. Figures~\ref{diphoton}--\ref{digluon} are the prediction of
our model.

From Fig.~\ref{diphoton}, we see that the diphoton signal strength can
be larger than unity when the dilaton decay constant $f$ is not much
larger than the SM Higgs vacuum expectation value (VEV), namely when
$\eta^{-1}=f/v\lesssim 1$. In a pure VBF/VH/ttH production
channel, only the negative~$\theta_H$ region can give an enhancement of the
diphoton signal strength. We can see from the right panels in
Figs.~\ref{diphoton}--\ref{digluon} that in purely dilatonic region
$\theta_H\simeq 0$, the VBF/VH/ttH production is suppressed for all the
decay modes. In particular, the other decay modes $WW, ZZ, bb,
\tau\tau$ etc.\ are always suppressed in the VBF/VH/ttH channel. The signal
strengths for decay modes other than $\gamma\gamma$ are generally
enhanced with GF production for a positive~$\theta_H$, as can be seen
in left panels of Figs.~\ref{others} and \ref{digluon}. Especially the
digluon signal strength can be enhanced as large as 30.

\subsection{Constraints on dilaton/Higgs sector}

\begin{figure}
\begin{center}
\hfill
\includegraphics[width=.3\textwidth]{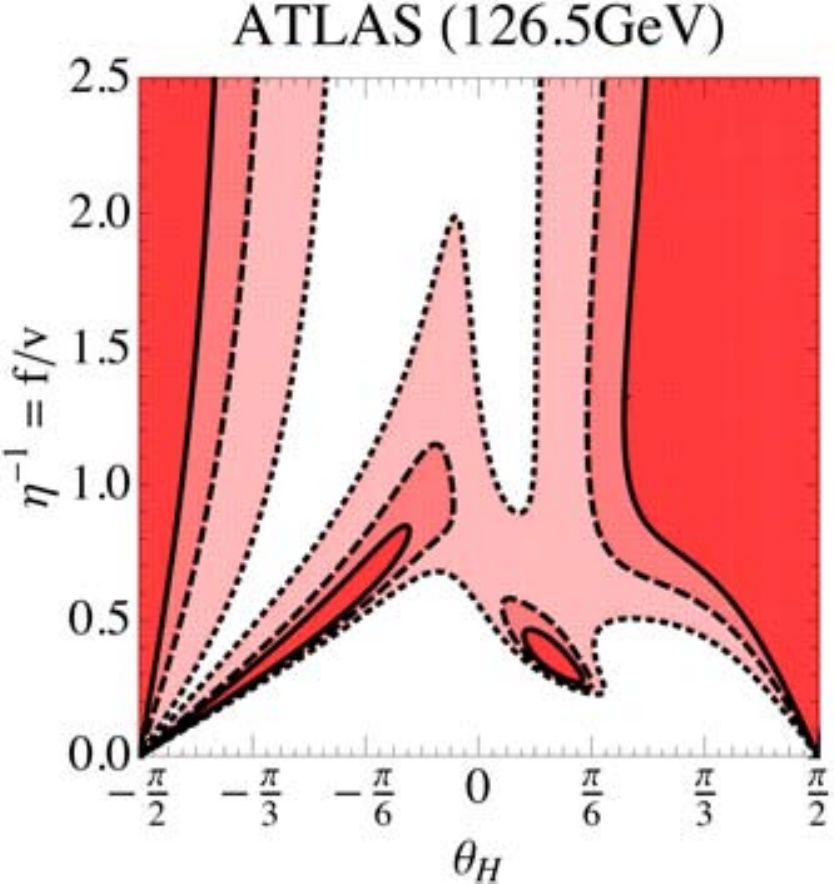}\hfill
\includegraphics[width=.3\textwidth]{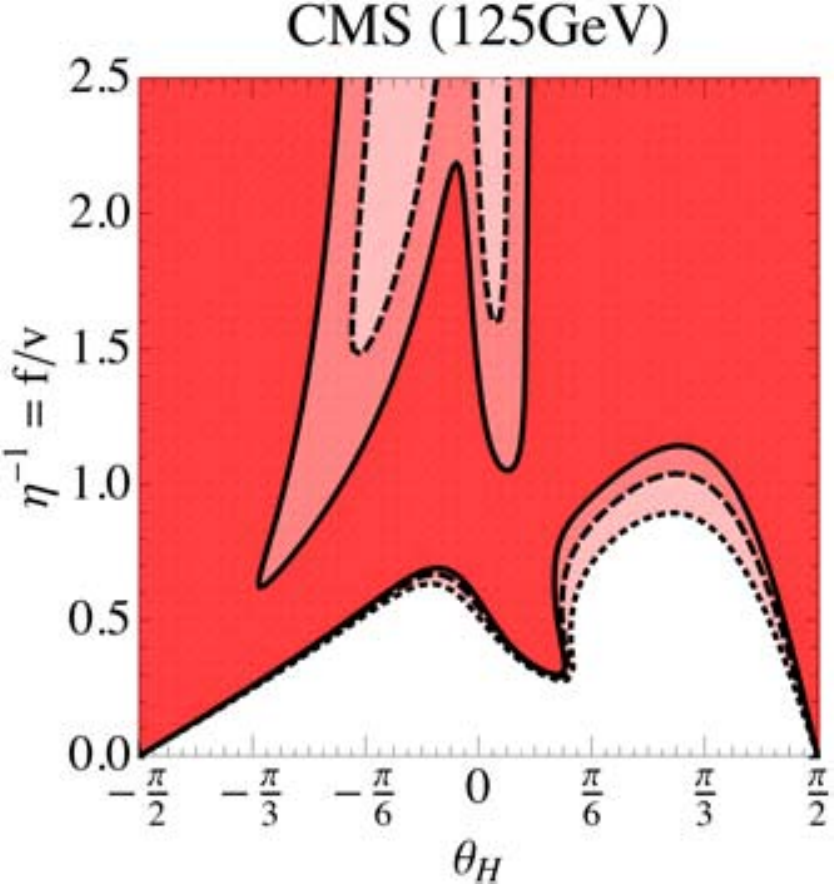}\hfill
\includegraphics[width=.3\textwidth]{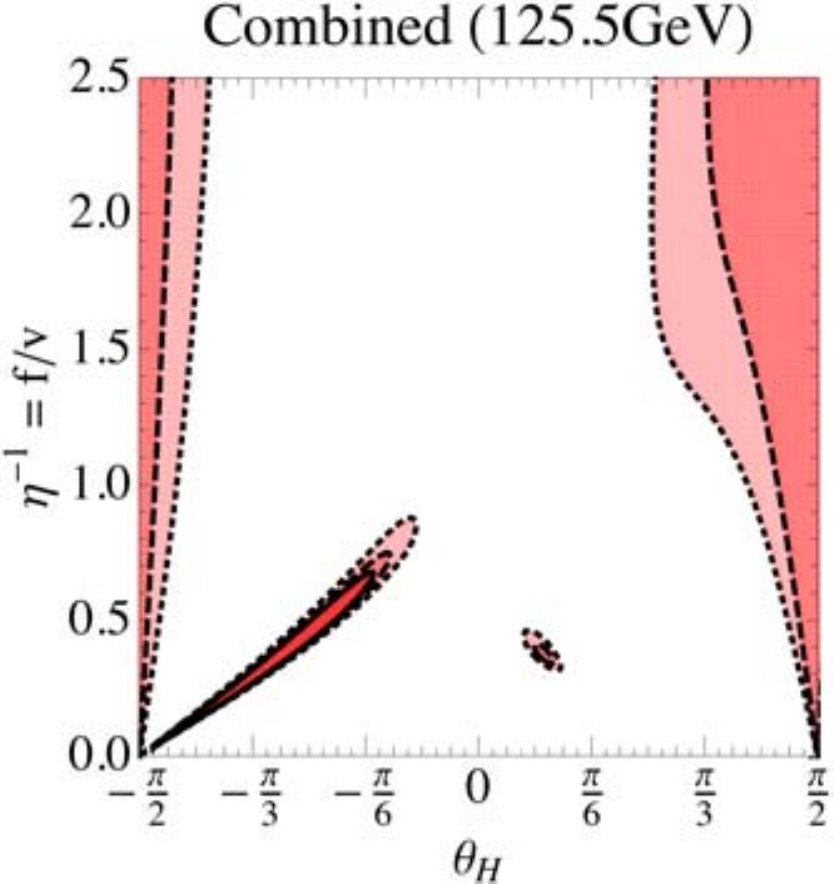}\hfill\mbox{}
\\
\caption{Favored regions within 90, 95 and 99\% confidence intervals, enclosed by solid, dashed, and dotted lines, respectively. Density (area) of favored region decreases (increases) in according order. Results are shown for ATLAS (left), CMS (center), and combined (right). See text for details.}
\label{constraints}
\end{center}
\end{figure}

As all the signal strengths are obtained, we perform a chi-square test with the Gaussian approximation for all the errors
\al{
\chi^2
	&=	\sum_i\paren{\muhat_i-\mu_i\over\sigma_i}^2,
}
where summation over $i$ is for all the diphoton categories as well as the $WW$ and $ZZ$ channels. For the $ZZ\to4l$ and $WW\to l\nu l\nu$ decay channels, we assume that all the signals are coming from GF and, hence, we approximate
\al{
\muhat(s\to VV)
	&=	\left|c_g\right|^2\,{R(s\to\text{others})\over R(s\to\text{all})}
}
for $VV=WW$ and $ZZ$. For ATLAS, the central value $\mu_i$ and deviation $\sigma_i$ are read off from Fig.~14 in Ref.~\cite{ATLAS_diphoton} for diphoton channels and from Fig.~10 in Ref.~\cite{obs:2012gk} for $WW$ and $ZZ$ channels. For CMS, Fig.~6(b) in Ref.~\cite{CMS_diphoton} and Table~7 in Ref.~\cite{obs:2012gu} are used for diphoton and $VV$ channels, respectively. The resultant number of degrees of freedom is 22 and 13 for ATLAS and CMS, respectively.

The results are shown in Fig.~\ref{constraints}. We see that both experiments have allowed dilatonlike region $\ab{\theta_H}<\pi/4$ within 90\% confidence interval though the ATLAS disfavors the purely dilatonic case $\theta_H\simeq0$ outside the 95\% confidence interval.
This is one of our main results.

As an illustration, we have also presented in right panel of Fig.~\ref{constraints} a ``theorist combination'' plot with the data from Fig.~3 in Ref.~\cite{Giardino:2012dp}. We have assumed that $WW$, $ZZ$, and $\gamma\gamma$ ($bbV$, $WWV$ and $\tau\tau$) are all coming from GF (VBF/VH/ttH) processes whereas $\gamma\gamma jj$ has 70\% from VBF/VH/ttH and 30\% from GF. In this naive treatment, we see that the SM is already outside the 90\% confidence interval whereas the minimal dilaton model has the allowed regions with a dilatonlike scalar.

\section{Constraints on top sector from electroweak data}\label{top_constraints}

When the Higgs-dilaton mixing $\theta_H$ is small, the relevant
parameters for the Peskin-Takeuchi $S$, $T$ parameters are the top partner mass $m_{t^\prime}$ and the left-handed top mixing $\theta_L$.
As we will see later, physics of the dilaton at the LHC is independent of
those two parameters $m_{t'}$, $\theta_L$ in the top sector. Therefore one can discuss the electroweak
constraints and the LHC physics independently. In this section, we
present allowed region of the parameters $m_{t^\prime}$ and $\theta_L$
from the electroweak precision measurements.

As is well known, the electroweak precision tests prefer a light Higgs
boson in the Standard Model. 
The upper bound is 185\,GeV at the 95\% confidence level (CL)~\cite{ALEPH:2010aa}.
On the other hand, the assumption that the 125\,GeV excesses at the LHC as
the dilaton requires the SM Higgs boson (if exists) to be
heavier than about 600\,GeV.
It is then necessary that the $t^\prime$ loops provide a correction
with an appropriate size and sign to push back to the allowed region in the $S$-$T$ plane.

We have obtained contributions to the Peskin-Takeuchi $S$, $T$ parameters from the $t$ and $t^\prime$ loops as
\begin{align}
 S^\text{top}
=&
\sin^2 \theta_L
\frac{N_c}{6 \pi}
\left[
\left(
\frac{1}{3}
-
\cos^2 \theta_L
\right)
\ln x
+
\left(
\frac{(1+x)^2}{(1-x)^2}
+
\frac{2 x^2 (3 - x ) }
{(1 - x)^3}
\ln x
-
\frac{8}{3}
\right)
\cos^2 \theta_L
\right]
, \label{S-top equation}
\\ 
 T^\text{top} 
=&
\sin^2 \theta_L
 \frac{N_c}{16 \pi} 
\frac{1}{s_W^2 c_W^2}
\frac{m_t^2}{m_Z^2}
\left[
\frac{\sin^2 \theta_L}{x}
-
(1 + \cos^2 \theta_L)
-
\frac{2}{1-x}
\cos^2 \theta_L
\ln x
\right]
,
\end{align}
where
\begin{align}
x := \frac{m_t^2}{m_{t'}^2}<1
\label{eq:def_of_x}
\end{align}
and $\theta_L$ is the mixing angle between $t$ and $t'$ defined in Appendix~\ref{top_mixing_section}.
If $\theta_L = 0$, then $t'$ decouples, and $S^\text{top}$ and $T^\text{top}$
become 0. This is because we have already subtracted the SM
contributions from the definition of $S$ and $T$ parameters, as usual.
For a fixed $\theta_L$ and a large $m_{t^\prime}$, $T^\text{top}$ is
enhanced as $\propto m_{t^\prime}^2$, whereas $S^\text{top}$ only has a logarithmic
dependence. Therefore, we generically obtain a large positive
contribution to $T^\text{top}$ and $|S^\text{top}| \ll T^\text{top}$.
Interestingly, this is indeed the required direction to come back to the
allowed region when the Higgs boson is heavy.

We also need to calculate the contributions from scalar sectors because
now we have two scalars, $s$ and $h$, and their couplings to the gauge
boson are different from the SM Higgs couplings.
We find 
\begin{align}
 S^\text{scalar}
=&
-
\frac{ \cos^2\theta_H }{12 \pi}
\ln \frac{m_\text{$h$ref}^2}{m_h^2}
-
\frac{ \sin^2\theta_H }{12 \pi}
\ln \frac{m_\text{$h$ref}^2}{m_s^2}
+
 \cos^2\theta_H f_S(m_h)
+
 \sin^2\theta_H f_S(m_s)
-
 f_S(m_\text{$h$ref})
\label{eq:S-scalar}
, \\ 
 T^\text{scalar}
=&
\frac{3 \cos^2\theta_H}{16 \pi c_W^2 }
\ln \frac{m_\text{$h$ref}^2}{m_h^2}
+
\frac{3 \sin^2\theta_H}{16 \pi c_W^2 }
\ln \frac{m_\text{$h$ref}^2}{m_s^2}
+
 \cos^2\theta_H f_T(m_h)
+
 \sin^2\theta_H f_T(m_s)
-
 f_T(m_\text{$h$ref})
\label{eq:T-scalar},
\end{align}
where $m_\text{$h$ref}$ is the reference Higgs boson mass
and 
$f_S(m_h)$ and $f_T(m_h)$
are small nonlogarithmic contributions whose explicit expression are  
\begin{align}
 f_S(m_h)
&=
-
\frac{1}{12 \pi}
\frac{(9 m_h^2 + m_Z^2) m_Z^4}{(m_h^2 - m_Z^2)^3}
\ln \frac{m_Z^2}{m_h^2}
-
\frac{(2 m_h^2 + 3 m_Z^2) m_Z^2}{6 \pi (m_h^2 -m_Z^2)^2}
, \\ 
 f_T(m_h)
&=
-
\frac{3 m_Z^2}{16 \pi s_W^2 c_W^2 (m_h^2 - m_Z^2)}
\ln \frac{m_h^2}{m_Z^2}
+
\frac{3 m_W^2}{16 \pi s_W^2 (m_h^2 - m_W^2)}
\ln \frac{m_h^2}{m_W^2}
.
\end{align}
Note that the $S$ and $T$ parameters given in~Eqs.~\eqref{eq:S-scalar} and
\eqref{eq:T-scalar} are independent of the sign of the Higgs-dilaton
mixing angle $\theta_H$ since they are functions of $\sin^2\theta_H$.

The contributions from $S^\text{scalar}$ and $T^\text{scalar}$ tend to
be smaller than the contributions from the 
top sector. The region in which they give non-negligible
contributions is around $\sin^2\theta_H =1$. In this region, we can not
ignore the sizable contribution from the term which is proportional to $\ln
(m_\text{$h$ref}^2/m_s^2)$. However,  
in this region, the $s$ couplings to the SM particles become almost the
same as the SM Higgs boson couplings, and $h$ behaves like SM singlet
particle. Then this region is nothing but the SM limit, which is not 
the interest in this paper. Therefore we can conclude that the dominant
contributions 
to the $S$ and $T$ parameters arise from the top sector, 
and $\theta_H$ dependence of the $S$ and $T$ parameters is mild.

\begin{figure}
\begin{center}
\hfill
\includegraphics[width=.3\textwidth]{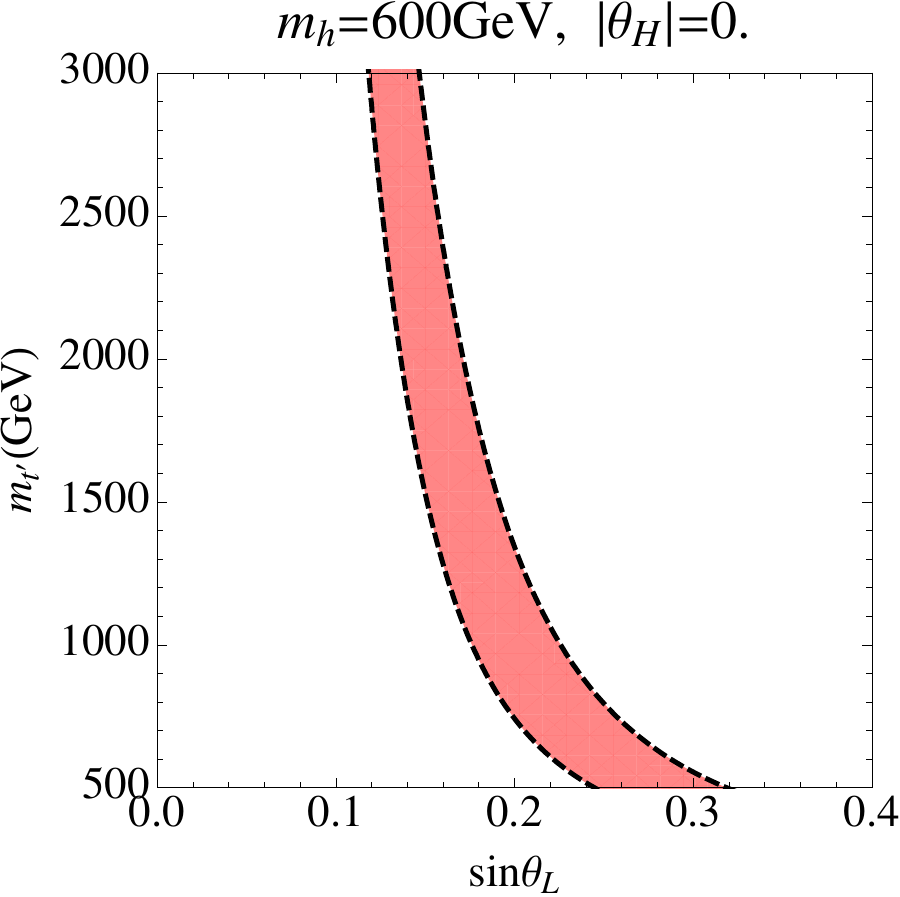}\hfill
\includegraphics[width=.3\textwidth]{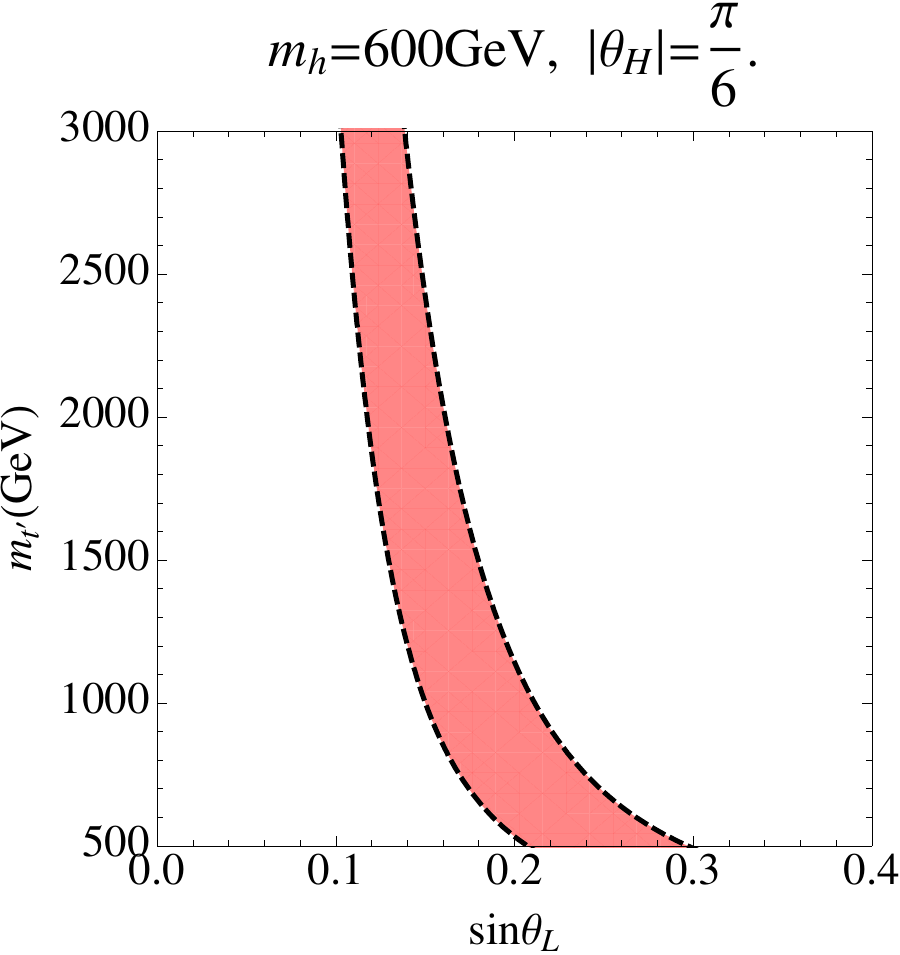}\hfill
\includegraphics[width=.3\textwidth]{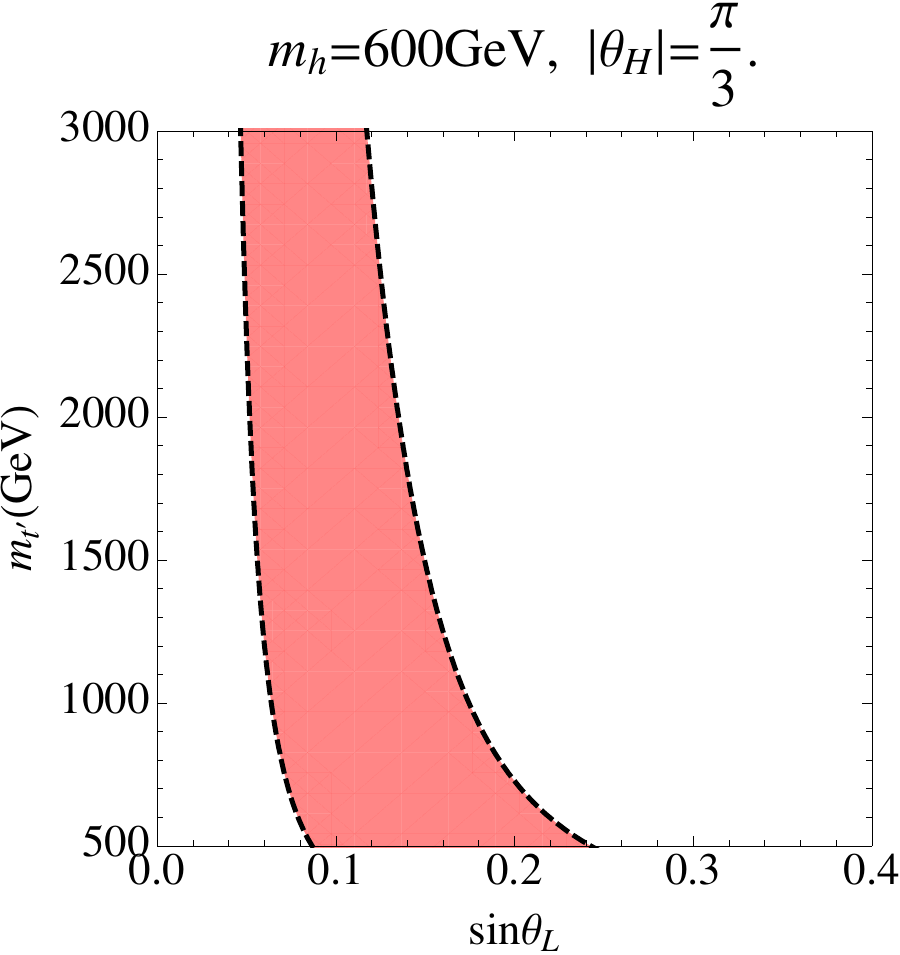}\hfill
\mbox{}\medskip\\
\hfill
\includegraphics[width=.3\textwidth]{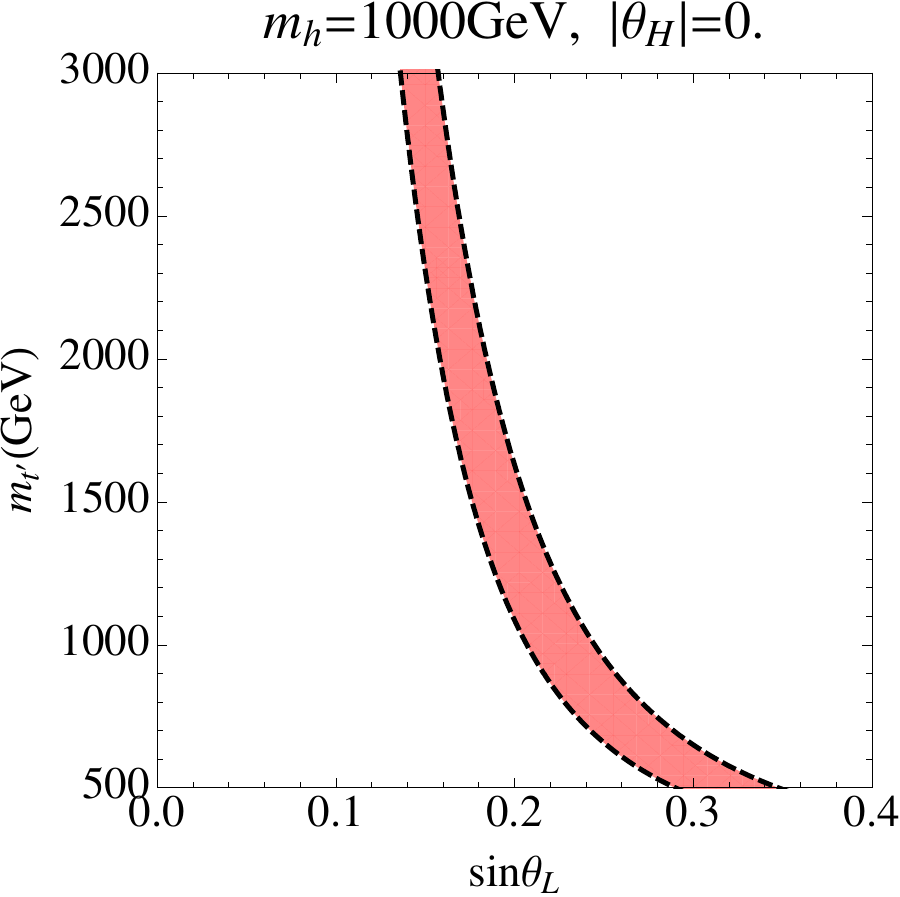}\hfill
\includegraphics[width=.3\textwidth]{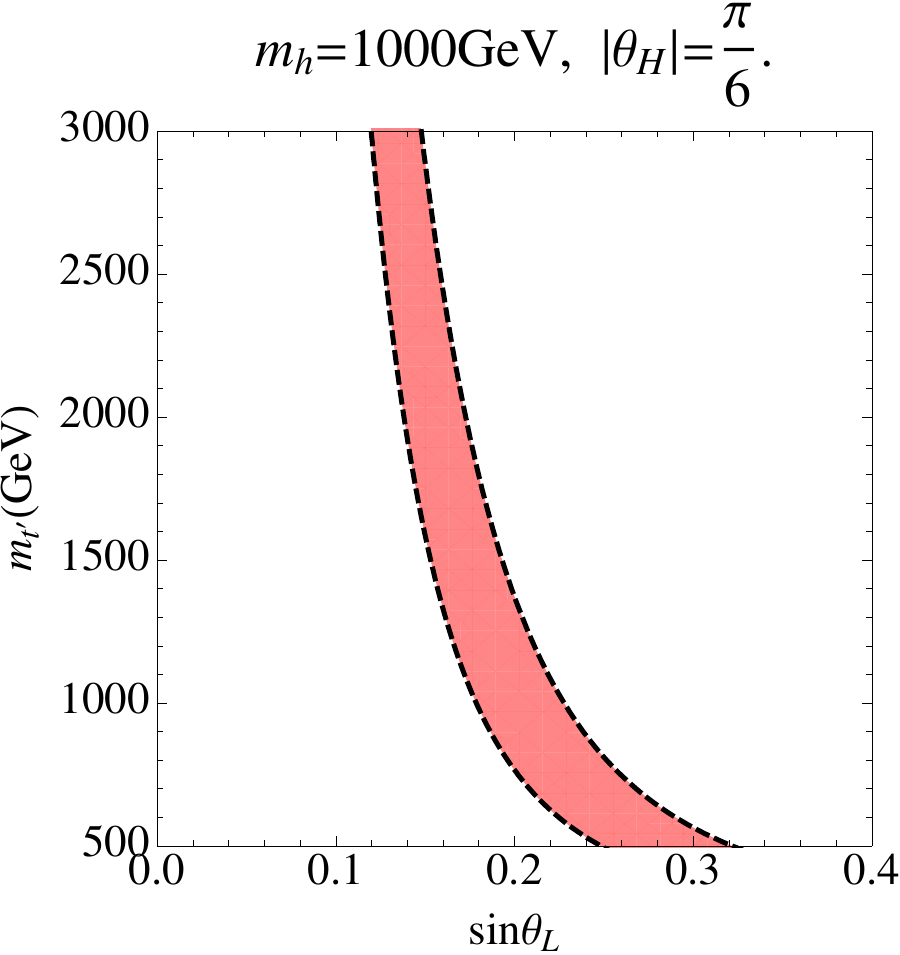}\hfill
\includegraphics[width=.3\textwidth]{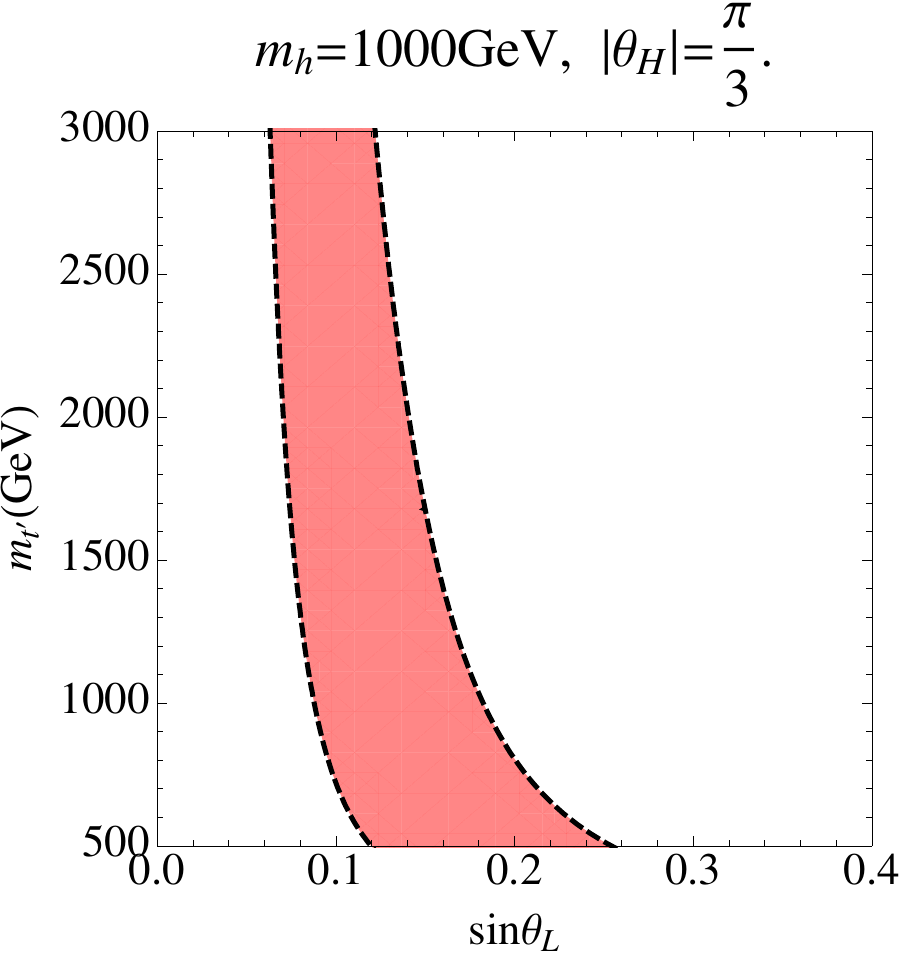}\hfill
\mbox{}
\\
\caption{Favored region plot from the Peskin-Takeuchi $S$, $T$. White regions are excluded at 95\% CL.}
\label{fig:ST}
\end{center}
\end{figure}

The numerical values of parameters we use are
\begin{align}
 s_W^2
 	&= 0.23, &
 v	&= 246\,{\rm GeV},	&
 m_s
 	&=	125\,{\rm GeV}, &
 m_h
 	&=	600,\,1000\,{\rm GeV},	&
\ab{\theta_H}
	&=	0,\,{\pi\over6},\,{\pi\over3}.
\end{align}
In Fig.~\ref{fig:ST}, we show favored region in the $m_{t^\prime}$ - $\sin \theta_L$ plane.
White regions are excluded at 95\% CL by $S$ and $T$ parameters.

There are other experimental constraints as well as $S$ and $T$ parameters.
The mass bound on $t^\prime$ from the direct search at the LHC
is~\cite{t' mass} 
\begin{eqnarray}
 m_{t^\prime} > 560~{\rm GeV}  ~~~~ (95 {\rm \% ~ CL})  .
\label{eq:mtp_exp_bound}
\end{eqnarray}
We can find a constraint on $\theta_L$ from the bound on $V_{tb}$ because
the mixing angle $\theta_L$ changes the top quark couplings, such as
$g_{Wtb}$. 
The bound
on $V_{tb}$ without assuming the unitarity triangle is~\cite{Abazov:2012vd}
\begin{align}
 0.81 < |V_{tb}| \leq 1 ~~~~ (95 {\rm \% ~ CL})  .
\end{align}
If we assume the top quark never mixed with light quarks, then the above
constraint gives
\begin{align}
 0.81 < | \cos \theta_L | \leq 1,
\end{align}
namely,
\begin{align}
 |\sin \theta_L| < 0.59.
\label{eq:bound_on_top_mixing}
\end{align}
We find that the constraints given in Eqs.~(\ref{eq:mtp_exp_bound}) and 
(\ref{eq:bound_on_top_mixing}) are easily satisfied in the allowed region in Fig.~\ref{fig:ST}.

We also study the constraint from $Zb_L b_L$ coupling. In the SM case,
the flavor-dependent corrections to this coupling are proportional to
the squared top-Yukawa coupling. In this model, this correction
is modified due to the mixing between $t$ and $t'$.
We parametrize $Z b_L \overline{b}_L$ coupling as follows
\begin{align}
 \frac{e}{s_W c_W}
\left(
-
\frac{1}{2}
+
 \delta g_L
+
\frac{1}{3}
s_W^2
\right).
\end{align}
We focus on only the flavor-dependent correction hereafter because we
use the constraint on~$R_b$~\cite{pdg:2012} to derive the constraint on
$Z b_L b_L$. 
We find that the flavor-dependent part of $\delta g_L$ is given by
\begin{align}
 \delta g_L 
&=
  \frac{m_t^2}{(4 \pi)^2 v^2}
+
 \delta g_L^\text{new}, 
\end{align}
where the first term is the SM contribution, and the second terms is the
additional contributions due to the $t'$ and the mixing angle
$\theta_L$.
We find 
\begin{align}
 \delta g_L^\text{new}
&= 
-
\frac{m_t^2}{16 \pi^2 v^2}
\frac{\sin^2 \theta_L}{( 1 - \sin^2 \theta_L (1 - x))^3}
\nn 
&
\qquad
\times
\Biggl(
-1 
-
\sin^4 \theta_L (1 - x)^3
+
2 x
+
\sin^2 \theta_L ( 2 - 5 x + 2 x^2)
+
\frac{
(1 - \sin^2 \theta_L)
x (1 + x) \ln x
}
{2 ( 1 - x)}
\Biggr)
\nn 
&\simeq
+
\frac{m_t^2}{16 \pi^2 v^2}
\frac{\sin^2 \theta_L}{( 1 - \sin^2 \theta_L )}
\qquad 
( x \ll 1 )
\label{eq:delta_gl_this_model}
,
\end{align}
where $x$ is defined in Eq.~(\ref{eq:def_of_x}).
The constraint on $\delta g_L^\text{new}$~\cite{Abe:2009ni}, 
which is derived from the constraint on $R_b$, 
is
\begin{align}
 \delta g_L^\text{new} 
&=
\left(
-5.8 \pm 8.6
\right)
\times
10^{-4}
.
\label{eq:Rb_const}
\end{align}
Comparing Eqs.~(\ref{eq:delta_gl_this_model}) and (\ref{eq:Rb_const}),
we find the region $
 |\sin \theta_L| > 0.52 
$
is excluded at 95\% CL.
This region is already excluded from the constraint on
$S$ and $T$ parameters in Fig.~\ref{fig:ST}.
Hence we conclude that the $Zb_L b_L$ constraint is not important in this model.

Before closing this section, we comment on the validity of our 
perturbative calculation. In the top sector, there are three parameters in the Lagrangian,
$m$, $y_t$, and $y'$. These three can be expressed by observables $m_t$, $m_{t'}$, and
$\theta_L$ as shown in Appendix~\ref{top_mixing_section}.
In the limit $m_{t'}\gg m_t$,
\al{
y_t	&\simeq
		{\sqrt{2}\over v}{m_t\over\cos^2\theta_L},
			&
y'	&\simeq
		{\sqrt{2}\over v}m_{t'}\sin\theta_L,	&
M	&\simeq
		m_{t'}\cos\theta_L.
}
We have seen that the small $\theta_L$ region is allowed by the $S$-$T$ bound. Taking $\theta_L\ll1$ limit, we get
\al{
y_t
	&\simeq
		{\sqrt{2}m_t\over v}
	\simeq
		1,	&
y'	&\simeq
		{m_{t'}\over m_t}\sin\theta_L,
		\label{eq:y'}	&
M	&\simeq
		m_{t'}.
}
We see that only $y'$ provides a nontrivial constraint, especially in
large $m_{t'}$ region. 
In Fig.~\ref{fig:ST}, we see that the allowed region of $\sin \theta_L$
is about less than 0.2 in large $m_{t'}$ limit.  Then, using
Eq.~(\ref{eq:y'}), we see that
\begin{align}
  y'
&\lesssim
 0.2\,\frac{m_{t'}}{m_{t}}.
\end{align}
From this relation, we find that if we impose that $y'$ should be lower than
4 to keep perturbativity, 
$m_{t'}$ should be lighter than 3400 GeV. This upper bound on $m_{t'}$
does not spoil our discussion in this paper.
Thus our perturbative calculation with $y'$ is valid unless we take
$m_{t^\prime}$ to be extremely large.\footnote{\addspan{
A tighter upper bound would be imposed, roughly $M\lesssim1$\,TeV depending on $\eta$, if one requires perturbativity for the coupling ${M\over f}S\overline{T}T$, though it is irrelevant for the computations in this paper.
}}

\section{Conclusions}

We have considered a possibility that the Higgs-like excesses observed
at the LHC experiments are actually the signals of the dilaton
associated with spontaneously broken scale invariance.
We have constructed a minimal model of the dilaton which can be produced
through the gluon fusion process at the LHC, and can decay into two
photons. The effective coupling is obtained through the loop diagrams of
a new vectorlike state $T$ that has the same gauge charges as the right-handed top quark.
The $T$ field contributes to the Peskin-Takeuchi $S$, $T$ parameters in the
electroweak precision tests. This contribution allows one to push the
Higgs boson mass above the experimental constraint, 600\,GeV,
providing a consistent framework for the light dilaton plus a heavy
Higgs scenario.

We find that the current experimental data allow distinct parameter
regions where the excesses are either Higgs-like or dilatonlike.
Once the excesses are confirmed with more statistics, it is possible to
distinguish two scenarios. \bigskip

\noindent Note added.---After completion of this manuscript, there appeared works treating a similar subject~\cite{note added}.

\subsection*{Acknowledgements}

The authors thank the Yukawa Institute for Theoretical Physics at Kyoto
University, where this work was initiated during the YITP workshop on
``Beyond the Standard Model Physics (YITP-W-11-27),'' March 19--23,
2012. The authors also acknowledge the participants of the workshop for
very active discussions.
We thank Masaya Ishino, Koji Nakamura, Yoshiko Ohno, \addspan{Masaharu Tanabashi}, and Junichi Tanaka for useful comments.
This work is supported in part
by the Grants-in-Aid for Scientific Research No.~23740165 (R.~K.), No.~23104009, No.~20244028, No.~23740192 (K.~O.), and No.~24340044 (J.~S.) of JSPS.

\appendix

\section{Mixing of top and its partner}\label{top_mixing_section}
As said in the text, we have chosen the basis on which the mass mixing between top and its partner $\oli{T_L}u_{3R}$ is rotated away
\al{
\bb
\oli{q_{3L}} & \oli{T_L}
\eb
		\bb
		m	&	m'\\
		0	&	M
		\eb
\bb
u_{3R} \\ T_R
\eb,
	\label{our choice}
}
where $m=y_tv/\sqrt{2}$ and $m'=y'v/\sqrt{2}$. Switching to mass eigenstates
\al{
\bb
q_{3L}\\
T_L
\eb
	&=	\bb
		\cos\theta_L	&	\sin\theta_L\\
		-\sin\theta_L	&	\cos\theta_L
		\eb
		\bb
		t_L\\
		t'_L
		\eb,	&
\bb
u_{3R}\\
T_R
\eb
	&=	\bb
		\cos\vartheta_R	&	\sin\vartheta_R\\
		-\sin\vartheta_R	&	\cos\vartheta_R
		\eb
		\bb
		t_R\\
		t'_R
		\eb,
}
we may diagonalize as
\al{
\bb
\oli{q_{3L}} & \oli{T_L}
\eb
		\bb
		m	&	m'\\
		0	&	M
		\eb
\bb
u_{3R} \\ T_R
\eb
	&=	\bb
		\oli{t_L} & \oli{t'_L}
		\eb
		\bb
		m_t	&	0\\
		0	&	m_{t'}
		\eb
		\bb
		t_R	\\ t'_R
		\eb,
}
where
\al{
\tan\theta_L
	&=	{\sqrt{\paren{M^2-m^2+m'^2}^2+4m'^2m^2}-M^2+m^2+m'^2\over2m'M}
	=	{m'\over M}+O(M^{-3}),	\nn
\tan\vartheta_R
	&=	{\sqrt{\paren{M^2-m^2+m'^2}^2+4m'^2m^2}-M^2+m^2-m'^2\over2m'm}
	=	{m'm\over M^2}+O(M^{-4}),
}
and the mass eigenvalues are
\al{
\left\{\begin{array}{c}m_t^2\\ m_{t'}^2\end{array}\right\}
	&=	
{M^2+m^2+m'^2\mp\sqrt{\paren{M^2+m^2+m'^2}^2-4m^2M^2}\over2}.
}
For large $M$,
\al{
m_t
	&=	\paren{1-{m'^2\over2M^2}}m+O(M^{-4}),\nn
m_{t'}
	&=	M+{m'^2\over2M}+O(M^{-3}).
}
Conversely, parameters in the Lagrangian can be written in terms of the observables:
\begin{align}
 M
&=
 \sqrt{ m_t^2 \sin^2 \theta_L + m_{t'}^2 \cos^2 \theta_L}
, \\ 
 y_t
&=
 \frac{\sqrt{2}}{v}
 \frac{m_t m_{t'}}{\sqrt{ m_t^2 \sin^2 \theta_L + m_{t'}^2 \cos^2
 \theta_L}}
, \\ 
 y'
&=
 \frac{\sqrt{2}}{v}
 \frac{(m_{t'}^2 - m_t^2) \sin \theta_L \cos \theta_L}{\sqrt{ m_t^2
 \sin^2 \theta_L + m_{t'}^2 \cos^2 \theta_L}} 
.
\end{align}

Instead of our choice~\eqref{our choice}, we may choose another basis where the Yukawa mixing $\oli{q_{3L}}T_R$ is erased
\al{
&	\bb
	\oli{q_{3L}} & \oli{T_L}
	\eb
	\bb
	m\cos\vartheta+m'\sin\vartheta	&	0\\
	M\sin\vartheta				&	M\cos\vartheta
	\eb
	\bb
	\widetilde u_{3R} \\ \widetilde T_R
	\eb,
}
where
\al{
\bb
\widetilde u_{3R} \\ \widetilde T_R
\eb
	&=
			\bb
			\cos\vartheta	&	\sin\vartheta\\
			-\sin\vartheta	&	\cos\vartheta
			\eb
	\bb
	u_{3R} \\ T_R
	\eb,	&
\tan\vartheta
	&=	{m'\over m}
	=	{y'\over y_t}.
}
Note that $\vartheta$ is not necessarily a small mixing angle and that the right-handed tops in this basis are related to the mass eigenstates by
\al{
\bb
t_R\\
t'_R
\eb
	&=	\bb
		\cos(\vartheta+\vartheta_R)	&	-\sin(\vartheta+\vartheta_R)\\
		\sin(\vartheta+\vartheta_R)	&	\cos(\vartheta+\vartheta_R)
		\eb
		\bb
		\widetilde u_{3R}\\
		\widetilde T_R
		\eb.
}
Although $\theta_L$ is more directly related to physical observables, one may trade it with $\vartheta+\vartheta_R$, which is the angle denoted by $\theta_u^R$ and is constrained in Ref.~\cite{Cacciapaglia:2011fx}.
We find the following relation between $\theta_L$ and
	$\vartheta+\vartheta_R \ (= \theta_u^R) $:
\begin{align}
 \sin \theta_L
=&
\frac{m_t}{\sqrt{m_t^2 \sin^2 \theta_u^R +  m_{t'}^2 \cos^2 \theta_u^R}}
 \sin \theta_u^R
.
\end{align}
Using this relation, we find our result is compatible with the result
given in 
Ref.~\cite{Cacciapaglia:2011fx}.

\section{Linear realization potential}\label{linear_realization_potential}
Though the precise form of the potential $\tilde V$ in Eq.~\eqref{linear_realization} is irrelevant for the experimental consequences which are governed by the Higgs-dilaton mixing angle~$\theta_H$ and the dilaton decay constant in units of the electroweak scale $\eta^{-1}=f/v$, let us write down a renormalizable linearized version of our potential just for completeness
\al{
\tilde V(S,H)
	&=	{m_S^2\over2}S^2
		+{\lambda_S\over4!}S^4
		+{\kappa\over2}S^2\ab{H}^2
		+m_H^2\ab{H}^2
		+{\lambda_H\over2^2}\ab{H}^4.
}
The VEVs for the SM Higgs $\left\langle
H\right\rangle=v/\sqrt{2}$ and for the singlet $\left\langle
S\right\rangle=f$ are obtained as 
\al{
\bb
f^2\\
v^2
\eb
	&=	{1\over{\lambda_S\lambda_H\over 6}-\kappa^2}
		\bb
		\lambda_H	&	-2\kappa\\
		-2\kappa	&	{2\lambda_S\over 3}
		\eb
		\bb
		-m_S^2\\
		-m_H^2
		\eb.
}
The mass eigenvalues are
\al{
\left\{\begin{matrix}m_s^2\\ m_h^2\end{matrix}\right\}
	&=	{\paren{{\lambda_S\over3}f^2+{\lambda_H\over2}v^2}
			\mp	\sqrt{
				\paren{{\lambda_S\over3}f^2+{\lambda_H\over2}v^2}^2
				-4\paren{{\lambda_S\lambda_H\over6}-\kappa^2}f^2v^2}
			\over
			2}
}
with the Higgs-dilaton mixing~\eqref{eq:thetaH} being
\al{
\tan2\theta_H
	&=	{2\kappa fv\over{\lambda_S\over3}f^2-{\lambda_H\over2}v^2}.
}
Since we want $m_s^2/m_h^2\lesssim\paren{125\GeV}^2/\paren{600\GeV}^2\simeq 4\%\ll1$, we can write
\al{
m_s^2
	&=	m_\sigma^2+O\fn{m_\sigma^4\over M_h^2},	&
m_h^2
	&=	M_h^2-m_\sigma^2+O\fn{m_\sigma^4\over M_h^2},
}
with
\al{
m_\sigma^2
	&:=	\paren{{\lambda_S\lambda_H\over6}-\kappa^2}{f^2v^2\over M_h^2},	&
M_h^2
	&:=	{\lambda_S\over3}f^2+{\lambda_H\over2}v^2,
}
where ${\lambda_S\lambda_H\over6}-\kappa^2>0$ is required in order not to have a tachyon.
Finally we note that a tree-level vacuum stability condition for a large VEVs reads ${2\lambda_S\over3}-4\kappa+\lambda_H>0$.

\end{document}